# Macroscopic 2D Wigner islands

M. Saint Jean[*], C. Even and C. Guthmann

Groupe de Physique des Solides, Universités Paris 6 / Paris 7
Unité mixte du C.N.R.S (U.M.R. 75 88),
2 place Jussieu, 75251 Paris Cedex ; France

**Abstract :**

In this paper we present new versatile «2D macroscopic Wigner islands» useful to investigate the various behaviors observed in mesoscopic confined systems. Our "Wigner islands" consist of electrostatically-interacting charged balls with millimetric size. We have experimentally determined the ground configurations for systems of N particles (N=1-30) confined in a parabolic potential and checked the influence of the confinement and interacting potentials. The results obtained are compared with the published numerical results.

**PACS :** 52.90.+z, 45.05.+x, 61.46.+w, 73.23.-b

# INTRODUCTION :

During the past few years many significant works have been dedicated to the study of the static and dynamic properties of mesoscopic systems. Among these studies, there is a particular stress on two important fields. The first one concerns the mesoscopic superconductors whose size is comparable to the coherence and penetration lengths and . The properties of the superconductor are determined by the value of the Ginzburg-Landau parameter = / . In type-II superconductors ( >1/ 2) the triangular Abrikosov vortex lattice is stable in the range $H_{c1}$<H<$H_{c2}$; however, in mesoscopic superconductors, the increase of the effective penetration length induces the appearance of an Abrikosov multivortex state even in type-I superconductors. Many papers describe the structure of a finite number of confined vortices and their transition from giant vortex [1]. The second up-to-date issue concerns the charging of 2D semiconductor quantum dots and the electrons coulomb blockade. Whereas it is commonly admitted that the minimum energy required to introduce one electron in a dot is $e^2$/C where C is the capacitance of the dot, it was observed that at low concentration of electrons, two electrons enter the dot simultaneously and that this particular effect repeats itself periodically with the number of electrons in the dot [2]. It has recently been recognized that this effect can be associated to the formation of highly symmetric electronic configurations [3].

All these mesoscopic systems of vortices or electrons can be assimilated to systems of a finite and small number of confined particles interacting with a Coulomb potential. Thus the knowledge of the particle configurations in these systems could be useful for understanding the properties of the mesoscopic devices. Therefore the determination of the ground or metastable state configurations, their excitation spectra or their phase transitions is a crucial issue. The systems of confined particles interacting by a Coulomb potential have been extensively studied during the past few years and many numerical works, involving Monte Carlo [4,5,6] or Molecular Dynamic [7,8] techniques, have been published on this topic. Unfortunately, although the reported patterns are qualitatively similar whatever the techniques used, the calculated ground state configurations sometimes differ.

On the other hand, various similar systems, like vortices in superfluid $He^4$ [9, 10], electron dimples on a liquid helium surface [11], trapped ions cooled by laser techniques [11], strongly coupled rf dusty plasma [12], were also experimentally studied. However, the results obtained are partial and differ, possibly because these experiments are complex and do not allow to simply tune the relevant parameters, for instance the number of interacting particles or the confinement potential.

In this paper, we present new versatile «2D macroscopic Wigner islands» which could be very convenient to investigate the various behaviors observed in mesoscopic confined systems. Our system consists in electrostatically-interacting charged balls of millimetric size moving on a plane conductor. The temperature is simulated by mechanical shaking. The main advantage of this system is that the number of particles is obviously perfectly controlled and that the confining potential can be tuned continuously without difficulty. Moreover the spatial configuration is directly observed at the macroscopic scale with a mere video camera. Let us also underline that the macroscopic scale of these islands allows to apply to them a controlled perturbative potential.

In section I, we describe the experimental setup, the procedures used to create the «2D macroscopic Wigner islands» and to reach the patterns corresponding to the ground state. We present the obtained arrays in section II whereas section III is devoted to the presentation of the numerical results and to their comparison with our experimental data.

**I - OVERVIEW OF THE SETUP**

The experimental setup is presented in figure 1. Our "Wigner islands" are constituted by metallic balls located in a horizontal plane capacitor to which a potential $V_e$ of about one thousand volts is applied. In order to confine the balls, another potential $V_c$ is applied to an isolated metallic frame located between the two capacitor electrodes. We used stainless steel balls of diameter $d = 0.8 mm$ and mass $m = 2.23 \pm 0.01 mg$. In order to eliminate possible contaminants on the surface of the balls, they are previously washed in a detergent, rinsed with water and ethanol, and dried in a drying-cupboard during several hours. The bottom capacitor electrode is a n-doped semiconductor wafer fixed on a metallic sheet, the top one being a transparent conducting glass. We chose such a wafer because it offers a perfect planar surface at the atomic scale; thus the friction force between the balls and the surface is considerably reduced. A metallic circular frame, isolated from these two electrodes, is intercalated between them. Its diameter is 10 mm, and its height can be chosen over a range of a few ball diameters. This experimental cell is fixed on a plate which leans on three loudspeakers.

The experiment is then the following one. Using the loudspeakers, the cell is initially strongly shaken and the system is liquid at this initial stage. The voltage $V_e$ is then applied instantanously between the capacitor electrodes, and is maintained throughout the experiment. The balls become monodispersely charged, repelling each other and spreading throughout the whole available space. We observe the existence of a voltage threshold of about 500 V below which the balls remain still. At this threshold, the repulsive electrostatic force acting between the balls is equal to the frictional force. Beyond this critical value, the charged balls move and interact freely. For the currently used $V_e$ potential, the charge of each ball is about $10^9$ electrons. Simultaneously the potential $V_C$ is applied to the external frame and the induced charges on this inter-electrode frame repel the charged balls. We emphasize that the advantage of this setup is to control independently the charge of the balls, their confining potential and the effective temperature.

In order to let the system explore its space of configurations and to find its minimum of energy, the cell is shaken using the three independent loudspeakers supplied by three white noises having a range $0-200\ Hz$. Shaking the balls using white noise produces random motion which simulates thermal Brownian motion corresponding to an effective temperature. The maximum frequency of $200\ Hz$ is chosen to be less than the fundamental modes of the different plates. This way of exciting the balls is efficient, since the minimum frequency for imparting their motion is $f_{min} = \sqrt{g/z_0}$, where $z_0$ is the vertical amplitude of displacement. In our experiment, $z_0 = 1$ mm, thus $f_{min} = 15$ Hz. In order to guarantee that the observed configurations correspond to the ground states with minimum energy, an annealing process is used. The system is initially heated up to its fusion temperature (in the sense of Lindeman [13]) and cooled down again at very small temperature. This first annealing is followed by a small number of additional ones, the effective temperature maximum (actually the loudspeaker amplitude) being at each step decreased by 10%. After these annealing cycles, the shaking is made at constant amplitude during a time varying from several minutes to several hours. Throughout the experiment, images of the arrays of balls are recorded in real-time using a CCD camera onto a VHS videocassette recorder. For a quantitative image analyzing, the camera is also connected to a computer containing an image grabber.

By analyzing these very long records and measuring the time spent in each observed configuration for a set of fixed experimental parameters, we retain for «ground state configuration» the most frequently observed state. Generally, this state is not different from the one obtained immediately after a sequence of annealing cycles, provided the number of cycles is higher than five.

## II. DESCRIPTION OF THE EXPERIMENTAL PATTERNS.

These original «macroscopic Wigner islands» (N<30) may be used as a very versatile model system to study the properties of a finite number N of charged particles interacting through a repulsive Coulomb potential and moving in a 2D parabolic confining potential. Indeed, the nature of the inter-ball force depends on the inter-electrode distance with respect to the ball dimension. For a large electrode distance, each ball can be considered as a point charge whereas the image charge in the top electrode has to be take into account in the case of a closer top electrode. Furthermore, the potential $V_C$ applied on the frame produces a confining potential with a cylindrical symmetry which creates a radial force varying with the distance from the cell center. Thus, even if the potential distribution $V_C(r)$ is determined by the frame height and its curvature radius, the confining potential $V_c(r)$ may be considered at first order as a parabolic potential. We will discuss later this point. We stress that in our setup, the intensity of the parabolic potential can be easely modified by changing the potential applied to the frame.

At low temperature, the observed arrays present self-organized patterns constituted by concentric shells on which the balls are located. In the following, we have called " state ($N_1$, $N_2$, $N_3$...)" the configuration in which $N_i$ is the occupation number in the ith shell from the center (Fig. 2). As it was widely discussed in the literature [4-12], this peculiar structure is due to the competition between the ordering into a triangular lattice symmetry which appears for infinite 2D coulombic system and the circular symmetry imposed by the confining potential. The strain energy being larger for bent structures, the effects of confinement are stronger for the smaller islands and thus especially important for the systems studied here (N<30).

The ground state configurations observed with our ball system are reported on Table 1. For N=3-5, regular polygons are formed. The first shell never exceeds five balls. For N=6, a centered pentagon (1,5) is formed. States with (1,6), (1,7), (1,8) and (2,8) patterns are obtained when N increases from 7 to 10. Further increasing N to 16 causes an alternate increase of $N_1$ and $N_2$ until both shells are full, respectively for 5 and 11. Notice that for these two shell structures, the second shell is intermediate between a circle and a polygon connecting the triangular lattice sites. For instance we can see that the (2,8) arrangement obtained for N=10 presents an elliptical shape due to the elongated center (fig 3). We can notice in figure 3 that a given shell configuration can correspond to different local order. Beyond N=16, a third shell appears; however unlike in a quantum atom in which new electrons are only present in the outer shell, this addition of balls induces an important reorganization of the pattern. Bedanov et al have explicited a construction rule which prescribes that "when all the shells are filled up to their maximum allowed numbers of particles, a new shell consisting of one ball is created in the center when one ball is added to the system". Our results validate this rule; for instance whereas the ground configuration for N=16 is (5,11), the ground configuration corresponding to N=17 is (1,5,11). Notice that in contrast with the two shells cases, the third outer shell is roughly circular whatever the inner core.

When we examine the ground configurations obtained when N increases, some of them appear at first glance more regular than the others (Fig.2). These islands with "magic numbers" of balls correspond to packings for which the difference between the occupation numbers $N_{i+1}-N_i$= 6 (i>1) in agreement with the hexagonal structure. In other words, these systems contain N=1+3p(p-1) balls, p being the number of shells. For these numbers, the system keeps the expected triangular lattice, the only change with respect to the ideal crystal being a slight adjustment near the outer shell.

For the non magic numbers, defects are present to accommodate the bending of the triangular lattice. In order to visualize how the circular outer shell and the inner triangular domain are adjusted and to compare the different configurations observed, it is convenient to describe the patterns with the commonly used -1 and +1 topological (or 5- and 7-fold) defects, corresponding to a ball with respectively 5 and 7 nearest neighbors. In the same framework, a ball located on the outer shell with three nearest neighbors has to be considered as a -1 topological defect since a defect-free site along a straight line lattice boundary has four nearest neighbors [8]. Following this analysis, each cluster with N>5 has -6 topological charges. For magic numbers systems, these charges correspond to the six corners of the hexagonal pattern and are all located in the outer shell, the inner part being defect-free. In contrast, for the other values of N, the defects move inside the structure. Our results show that when different configurations are observed for a given N, the ground state is always the pattern for which the defects are located in the outer shell. For instance, for N=27, our ground state is the configuration (3,9,15) which is very close to a magic number system whereas the first metastable configuration (4,9,14) presents a 5-fold defect inside. Let us indicate that this result is in complete disagreement with the opposite analysis given by Ying-Ju Lai et al in reference 8.

Metastable configurations are observed at the same time as the ground state configurations. Figure 4 shows for instance the two configurations observed for N=17. This case is very spectacular since it is at the border between the two and three shells structures : the ground configuration presents three shells whereas the first metastable state is a two shells structure. Another interesting case is N=6. For this number, we observed a ground configuration (1,5) and the metastable state corresponding to the hexagonal configuration (Fig.4). Moreover, the unstable equilibrium associated with a saddle point in the configuration space is also observed; it corresponds to the transition between the (1,5) and the hexagonal configurations. In this state the angle between two balls in the outer shell is larger than $2\pi/5$, the inner ball being out of the central position. These three configurations have been calculated by Bolton et al [5], the inner ball excentricity being evaluated to $0.6 x_0$ where $x_0$ is the pentagone radius. From our data, we obtained a smaller deviation of the central ball, about $0.2 x_0$, for this unstable equilibrium.

In order to evaluate the influence of the confinement potential on the ground configurations observed, we varied the $V_c$ potential applied to the frame. No modifications in the obtained ground configurations were observed when the $V_c$ potential was increased by a factor of 2 from 600 to 1200V, the bottom electrode being maintained at 600V. We also checked the influence of the interacting potential. By modifying the balls-transparent electrode distance over a few ball diameters, we modified the range of the inter-balls interaction. Indeed, when the electrode is far from the balls the interaction is a purely Coulomb one whereas, because of the existence of image charges, this interaction becomes dipolar when the electrode is closer. Once more, the observed ground configurations are similar whatever the electrode distance.

## III DISCUSSION

Rare analytical studies have been performed on such systems of confined particles in Coulomb interaction. In contrast, an important numerical activity, involving both Monte Carlo and Molecular Dynamic methods, has been developed. Depending on the cases, these studies determined the ground configurations [4-8], the metastable states [4,8], the excitation spectra [7,8], the phase transitions [6], or the topological defects [8] of these confined systems. To be exhaustive, we have to indicate that few studies introduced different interaction potentials and different shapes of confining potential in order to evaluate their influences on the ground configurations [6, 8]. In table I, we compile the different ground states obtained in these studies. We report the results obtained by Monte Carlo (MC) and Molecular Dynamic (MD) methods. All these configurations correspond to a system of particles in Coulomb interaction confined in a parabolic potential.

First, our results shed some light on two assumptions proposed in the literature concerning the influence of the confining or interacting potentials. In order to discuss these points, let us consider the three shells structures for which one can expect these effects to be relevant. As it will be shown later, our ground state configurations are in complete agreement with those obtained by Campbell et all for a parabolic confinement. Furthermore, these configurations are experimentally independent of the confining potential when the latter is increased by a factor of 2. On the other hand, Ying-Ju et al have tested by MD calculations the influence of a steep confining potential on the ground states configurations. They showed that, even if the shell structures are still obtained, the stiffening of the confinement results in an increase of the density of the outer shell [8]. The results obtained by Bedanov et al and for a hard wall confinement [6] are qualitatively discussed in the same way. Our experiments do not confirm this dependence. Moreover, our own results suggest that the confinement effect is not the sole reason for the increase of the outer shell densities. Indeed, the packing densities of the outer shells in our configurations are generally higher than those obtained by the numerical methods described in references 6 and 8 for a parabolic confinement, but remain smaller than the corresponding ones calculated in a strong confinement potential. Another comment concerning the influence of the inter-balls interaction range suggested that the short range of a screened interaction potential facilitates a higher packing density in the inner core [8]. Our data do not confirm these numerical results. Indeed, our ground configurations were identical when we modified the balls-top electrode distance over a few ball diameters; in other words, the observed results are independent of the range of the inter-balls interaction.

The main interest of our experiment is to allow the selection among the numerical methods used to describe the behaviors of systems of confined Coulomb-interacting particles .Indeed, although shell structures are obtained whatever the numerical methods used, many differences can be observed in their identification of the ground state configurations.

For very small islands with one or two shells (N<16), the ground state configurations are the same whatever the numerical method used, except for the noticeable cases N=9 and 15. For the latter cases, the configurations generally retained as ground states are respectively (2,7) and (5,10) whereas Campbell and Zip [4] using MC method proposed (1,8) and (4,11) configurations (hereafter named CZ results) in agreement with our experimental results and the ones obtained in reference 11. This originality of the CZ results can also be observed for N=16 for which they proposed a two shells (5,11) structure whereas all the other groups proposed a three shells pattern (1,5,10). For N>17, all the calculated patterns present three shells, however the situation is much more confused than previously. Even if the two MD studies report the same results (except for N=29), in contrast the various MC calculations present many differences between themselves. This last point is particularly obvious if we compare for instance the ground state configurations obtained respectively by Campbell et al [4] and Bolton et al [5], almost all the calculated configurations are different. In contrast, the MC calculations performed by Bedanov et al are in almost complete agreement with those of Bolton et al [5]. Moreover, Schweigert and Peeters indicate in reference 7 that their MD results are in complete agreement with the Bedanov and Peeters MC calculations [6]. Thus, once more, the CZ results seem singular. Beyond N=26, this originality of the CZ results cannot be estimated since these authors did not calculate the corresponding configurations. However, all the other numerical results are in agreement, except the (4,10,15) configuration proposed by Ying-ju et al [8] which differs from the (5,10,14) proposed by the other groups.

In fact, these apparently very large discrepancies between the numerical results have to be analyzed more carefully. The cases where different ground states are determined can be analyzed using the energies calculated for the ground and metastable configurations by Campbell et al (who are alone to give exhaustively this kind of data). In all these cases, the ground configurations calculated by the other groups always correspond to a CZ metastable states. This state is generally the nearest from the CZ ground state but not necessarily. For instance the (1,6,10) and (3,9,13) configurations proposed by the other groups corresponds respectively to the fourth and fifth metastable states obtained by Campbell et al for N=17 and 25. Let us indicate that for N=22 Campbell et al did not report the (2,8,12) configuration which is probably at higher energy for them. However three different situations can be identified depending on the relative energy differences $\Delta E/E$ between the Campbell et al ground state and its metastable state which is recognized as ground state by the other groups. Many non identical ground configurations present strong relative energy difference $\Delta E/E>30\%$; it is the case for N= 9, 16, 17, 22, 24, 25 ; for N=15 and 23, $\Delta E/E$ is about 10%:

in contrast only one case, N= 20, presents a very small relative energy difference ($\Delta E/E < 10^{-5}$).

Thus from this analysis of the relative energy difference $\Delta E/E$, we can identify the cases for which the differences between the CZ results and the other ones are significant. For small two shells systems, it seems that the confinement bending is not strong enough to induce a great variety of configurations; then the different states are well identified and all the calculated ground states are identical whatever the numerical method used. Our experiments confirm the calculated patterns. The particular case N=9 is discussed below. As N increases, the confinement is more efficient and the patterns stand more metastable states. This increase of the number of metastable states induced specific difficulties in numerical studies which can partially explain the observed discrepancies. Indeed, if we consider the systems N= 15, 20 and 23 for which $\Delta E/E$ is small, the metastable states are nearly degenerated with the ground state. The diversity of the retained ground state configurations could be interpreted as resulting from numerical error to determine in a finite time the minimum energy, as soon as there exists a lot of local minimum configurations with energies very close to the ground energy. On the contrary, for N= 9, 16, 17, 22, 24, 25, this difference is now relevant since in these cases, $\Delta E/E$ is large, and can result from the intrinsic difference in the numerical methods used.

By comparing our experimental results with numerical ones, we observe that our own patterns are in excellent agreement with Campbell's ones, except for N= 23 for which we obtain the configuration (2,8,13) proposed by the other groups, but this latter case corresponds to an irrelevant misfit since $\Delta E/E$ is small. Thus our experimental results unambiguously confirm the ground configurations solely calculated by Campbell et al. Therefore it seems that the Campbell calculations offer the best opportunities to evaluate the ground states, yet we cannot presently explain the reasons of this advantage.

In summary, we have demonstrated a simple experimental setup convenient for studying the properties of systems constituted by equally charged particles in coulombic interaction. If sufficient experimental care is taken, the systems with a small number of balls are equivalent to superconductor or semiconductor mesoscopic systems containing few interacting vortices or electrons. Moreover, our results are in complete agreement with Campbell and Zip results, which shows the complementary use of our system with respect to numerical approaches.

**TABLE I**

| N | Our results | Campbell et al (MC) | Bolton et al (MC) | Bedanov et al (MC) | Schweigert et al (MD) | Ying-ju et al (MD) |
|---|---|---|---|---|---|---|
| 1 | 1 | 1 | 1 | 1 | 1 | |
| 2 | 2 | 2 | 2 | 2 | 2 | |
| 3 | 3 | 3 | 3 | 3 | 3 | 3 |
| 4 | 4 | 4 | 4 | 4 | 4 | 4 |
| 5 | 5 | 5 | 5 | 5 | 5 | 5 |
| 6 | 1/5 | 1/5 | 1/5 | 1/5 | 1/5 | 1/5 |
| 7 | 1/6 | 1/6 | 1/6 | 1/6 | 1/6 | 1/6 |
| 8 | 1/7 | 1/7 | 1/7 | 1/7 | 1/7 | 1/7 |
| 9 | 1/8 | 1/8 | 2/7 | 2/7 | 2/7 | 2/7 |
| 10 | 2/8 | 2/8 | 2/8 | 2/8 | 2/8 | 2/8 |
| 11 | 3/8 | 3/8 | 3/8 | 3/8 | 3/8 | 3/8 |
| 12 | 3/9 | 3/9 | 3/9 | 3/9 | 3/9 | 3/9 |
| 13 | 4/9 | 4/9 | 4/9 | 4/9 | 4/9 | 4/9 |
| 14 | 4/10 | 4/10 | 4/10 | 4/10 | 4/10 | 4/10 |
| 15 | 4/11 | 4/11 | 5/10 | 5/10 | 5/10 | 5/10 |
| 16 | 5/11 | 5/11 | 1/5/10 | 1/5/10 | 1/5/10 | 1/5/10 |
| 17 | 1/5/11 | 1/5/11 | 1/5/11 | 1/6/10 | 1/6/10 | 1/6/10 |
| 18 | 1/6/11 | 1/6/11 | 1/6/11 | 1/6/11 | 1/6/11 | 1/6/11 |
| 19 | 1/6/12 | 1/6/12 | 1/6/12 | 1/6/12 | 1/6/12 | 1/6/12 |
| 20 | 1/6/13 | 1/6/13 | 1/7/12 | 1/7/12 | 1/7/12 | 1/7/12 |
| 21 | 1/7/13 | 1/7/13 | 2/7/12 | 1/7/13 | 1/7/13 | 1/7/13 |
| 22 | 1/7/14 | 1/7/14 | 2/8/12 | 2/8/12 | 2/8/12 | 2/8/12 |
| 23 | 2/8/13 | 1/8/14 | 2/8/13 | 2/8/13 | 2/8/13 | 2/8/13 |
| 24 | 2/8/14 | 2/8/14 | 3/8/13 | 3/8/13 | 3/8/13 | 3/8/13 |
| 25 | 3/8/14 | 3/8/14 | 3/9/13 | 3/9/13 | 3/9/13 | 3/9/13 |
| 26 | 3/9/14 | 3/9/14 | 3/9/14 | 3/9/14 | 3/9/14 | 3/9/14 |
| 27 | 3/9/15 | | 4/9/14 | 4/9/14 | 4/9/14 | 4/9/14 |
| 28 | 3/9/16 | | 4/10/14 | 4/10/14 | 4/10/14 | 4/10/14 |
| 29 | 4/9/16 | | 5/10/14 | 5/10/14 | 5/10/14 | 4/10/15 |
| 30 | 4/9/17 | | 5/10/15 | 5/10/15 | 5/10/15 | 5/10/15 |

**References**

1: V. A. Schweigert, F. M. Peeters, P. Singha Deo, Phys. Rev. Lett. **81**, 2783 (1998) and references therein.

2: N.B Zhitenev, R.C. Ashoori, I. N. Pfeiffer, K. W. West, Cond-Mat 9703241.
n
3: A. A. Koulakov, B. I. Shklovskii, Cod-Mat 9705030

4: L.J. Campbell, R.M. Ziff, Phys. Rev. B **20**, 1886 (1979).

5: F. Bolton, U. Rössler, Superlattices and Microstructures **13**, 139 (1993).

6: V. Bedanov, F. M. Peeters, Phys.Rev.B **49**, 2667 (1994).

7: V. A. Schweigert, F. M. Peeters, Phys. Rev. B **51**, 7700 (1995).

8: Ying-Ju Lai, Lin I, Phys. Rev. E **60**, 4743 (1999).

9: E. J. Yarmachuk, R. E. Packard, J. Low Temp. Phys. **46**, 479 (1982).

10: W. I. Galberson, K. W. Schwarz, Physics Today **40**,54 (1987).

11: P. Leiderer, W. Ebner, V. B. Shikin, Surf. Sci. **113**, 405 (1987).

12: Wen Tau Juan, Zen-hong Huang, Ju-Wang hsui, Ying-Ju Lai, Lin I, Phys. Rev. E **58**, R6947 (1998).

13: F. Lindeman, Z. Phys. **11**, 609 (1910).

14: L. Bonsall and A. A. Maradudin, Phys. Rev. B **15**,. 1959 (1977)

**Figure captions**

Fig 1 : Setup overview.

Fig. 2 : Examples of ground state configurations observed for N= 5 to 29. The diameter of the circular frame is 10 mm.

Fig.3 : Two configurations recorded for N=10. During the shaking, the inner shell moves very fast between these two positions suggesting that these two "ground"configurations are very close in energy. Notice the elliptical shape of the second shell.

Fig. 4 : Ground and metastable configurations observed for N=17 and 6. Patterns observed for N=6; a) the ground state configuration (1,5); b) the hexagonal metastable configuration; c) the instable equilibrium corresponding to the saddle point in the configuration space to transit from (1,5) to the hexagonal shape.

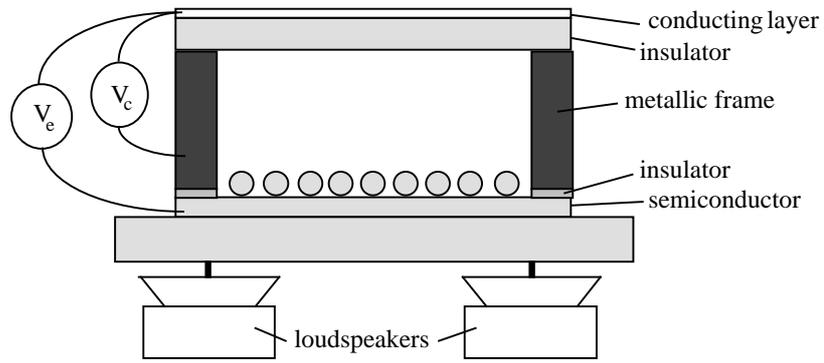

Fig. 1

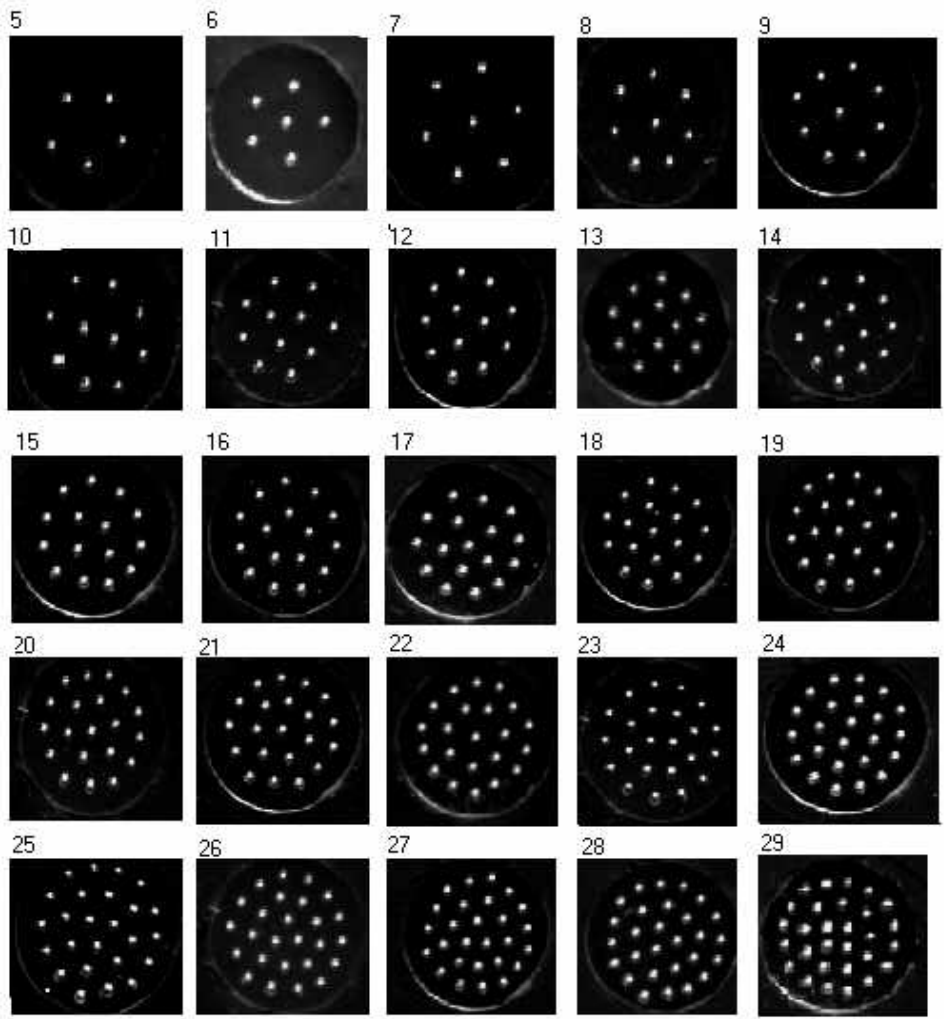

Fig. 2

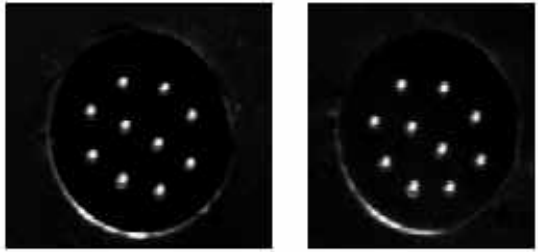

Fig. 3

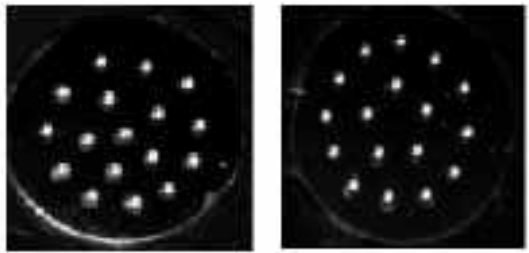

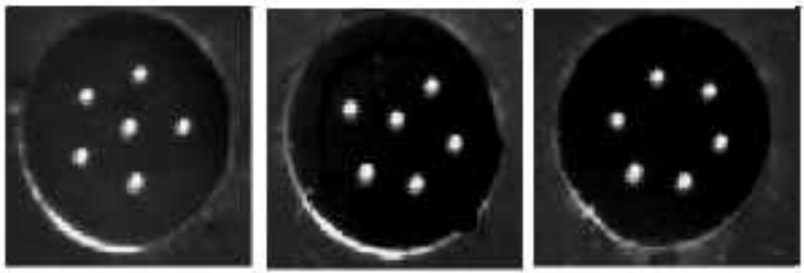

Fig. 4